\documentclass[letterpaper, 10pt, conference]{ieeeconf}

\IEEEoverridecommandlockouts        
\usepackage{cite}
\usepackage{hyperref}
\usepackage{balance}
\usepackage{blkarray}
\usepackage{multirow}
\usepackage{booktabs}

\newcounter{definition}
\newtheorem{theorem}{Theorem}
\newtheorem{lemma}[theorem]{Lemma}
\newtheorem{corollary}[theorem]{Corollary}
\newtheorem{proposition}[theorem]{Proposition}

\newtheorem{remark}{Remark}

\newcounter{definitionctr}

\renewcommand{\thedefinitionctr}{Definition~\arabic{definitionctr}}

\newenvironment{definition}[1][]{
    \refstepcounter{definitionctr} 
    \noindent\textit{\thedefinitionctr. #1} 
}

\def\pfof[#1]{\noindent\hspace{2em}{\itshape Proof of #1: }}

\usepackage{graphicx}      
\usepackage{amsmath}
\usepackage{amssymb}
\usepackage{xcolor}
\usepackage{subfigure}
\usepackage{comment}
\usepackage{tikz}

\def\R{\mathbb{R}}
\def\C{\mathbb{C}}
\def\G{\mathcal{G}}
\def\V{\mathcal{V}}
\def\E{\mathcal{E}}
\def\P{\mathcal{P}}

\newcommand{\diag}{\mathrm{diag}}
\newcommand{\rank}{\mathrm{rank}}
\newcommand{\srank}{\mathrm{s}\text{-}\mathrm{rank}}

\newcommand\oprocendsymbol{\hbox{$\square$}}
\newcommand\oprocend{\relax\ifmmode\else\unskip\hfill\fi\oprocendsymbol}

\definecolor{aoenglish}{rgb}{0.0, 0.5, 0.0}
\definecolor{darkblue}{rgb}{0.0, 0.0, 0.55}
\definecolor{darkmagenta}{rgb}{0.55, 0.0, 0.55}
\definecolor{electricviolet}{rgb}{0.56, 0.0, 1.0}
\definecolor{electricyellow}{rgb}{1.0, 1.0, 0.0}
\definecolor{forestgreen}{rgb}{0.13, 0.55, 0.13}
\definecolor{fuchsia}{rgb}{1.0, 0.0, 1.0}
\definecolor{gamboge}{rgb}{0.89, 0.61, 0.06}
\definecolor{goldenpoppy}{rgb}{0.99, 0.76, 0.0}
\definecolor{indigo}{rgb}{0.29, 0.0, 0.51}
\definecolor{internationalorange}{rgb}{1.0, 0.31, 0.0}
\definecolor{lava}{rgb}{0.81, 0.06, 0.13}
\definecolor{selectiveyellow}{rgb}{1.0, 0.73, 0.0}
\definecolor{turquoiseblue}{rgb}{0.0, 1.0, 0.94}
\definecolor{turquoise}{rgb}{0.19, 0.84, 0.78}

 \title{\LARGE \bf How Complex is a Complex Network? \\ Insights from Linear Systems Theory}

\author{Giacomo Baggio, Marco Fabris
\thanks{This work was partially carried out within the Italian National Center for Sustainable Mobility and received funding from NextGenerationEU (Italian NRRP – CN00000023 - D.D. 1033 17/06/2022 - CUP C93C22002750006).}
\thanks{The authors are with the Dipartimento di Ingegneria dell'Informazione, Universit\`a di Padova, Via Gradenigo 6/B, 35131 Padova, Italy. E-mails: \href{mailto:giacomo.baggio@unipd.it}{\texttt{giacomo.baggio@unipd.it}}, \href{mailto:marco.fabris.1@unipd.it}{\texttt{marco.fabris.1@unipd.it}}.}
}

\usepackage{graphicx,color}
\usepackage{epstopdf}
\usepackage{amsmath}
\usepackage{amssymb}
\usepackage{mathrsfs}
\usepackage{subfigure}
\usepackage{url}
\usepackage{color}
\begin{document}
\maketitle

\thispagestyle{empty}
\pagestyle{empty}

{\begin{abstract}
This paper leverages linear systems theory to propose a principled measure of complexity for network systems. We focus on a network of first-order scalar linear systems interconnected through a directed graph. By locally filtering out the effect of nodal dynamics in the interconnected system, we propose a new quantitative index of network complexity rooted in the notion of McMillan degree of a linear system. First, we show that network systems with the same interconnection structure share the same complexity index for almost all choices of their interconnection weights. Then, we investigate the dependence of the proposed index on the topology of the network and the pattern of heterogeneity of the nodal dynamics. Specifically, we find that the index depends on the matching number of subgraphs identified by nodal dynamics of different nature, highlighting the joint impact of network architecture and component diversity on overall system complexity.
\end{abstract}}


\section{Introduction}


The analysis of complex networks is central to various scientific and engineering disciplines, ranging from power grids and biological networks to social systems and communication infrastructures.  In many of these areas, the nodes that form the network are not static entities but dynamic systems that are interconnected according to prescribed patterns. To effectively quantify the complexity of such network systems, it is crucial to consider not only the (static) pattern of interactions among nodes but also the~dynamic~nature~of~the~nodes. 

A range of indices have been proposed to quantify complexity in networks, drawing from different theoretical frameworks \cite{costa2007characterization,kim2008complex,harrison2016role}. Some metrics focus on the computational complexity of performing numerical operations on the adjacency matrix of the network, such as matrix-vector multiplication \cite{neel2013linear}. Others use structural features, like the number of edges, connected components, or number of spanning trees, as proxies for complexity \cite{minoli1975combinatorial,grone1988bound,bonchev2005quantitative}. Structural heterogeneity is also captured through metrics that count distinct (non-isomorphic) subgraphs within the network \cite{kim2008complex} or entropy-based measures applied to correlation matrices characterizing the graph structure \cite{claussen2007offdiagonal}. 
While designed for static networks, such metrics have been often applied to networks of~interconnected~dynamical~systems.

Several studies have proposed measures of complexity of dynamical systems in terms of factors such as the ability of the system to produce diverse behaviors, unpredictability of trajectories, sensitivity to initial conditions, or robustness-fragility trade-offs, see e.g.~\cite{lloyd2001measures,carlson2002complexity,ladyman2013complex}. When restricting to linear time-invariant systems, a simple and canonical measure of complexity is the McMillan degree, representing the minimal dimension of a state-space realization that is both controllable and observable \cite{kalman1965irreducible}. To our knowledge, there are no existing metrics for interconnected systems that account for the complexity arising from interconnection patterns, even in the case of linear time-invariant systems.

In this paper, we explore a simple yet insightful scenario involving network systems composed of interconnected first-order linear dynamical nodes and introduce a network complexity index inspired by an input reconstruction problem with local information. Specifically, we apply a local deconvolution step to the state of the network system to locally filter out the effect of nodal dynamics in the reconstruction process. We then define network complexity as the McMillan degree of the resulting system, which quantifies the residual complexity of reconstructing external inputs after removing local node dynamics. We show that the proposed complexity index depends almost exclusively on the interconnection pattern of the system, as it takes the same value for almost all choices of the interconnection weights. We then clarify the dependence of the proposed index on the interconnection structure and the heterogeneity of the nodal dynamics by establishing a characterization in terms of the matching numbers of subgraphs corresponding~to~different~local~dynamics.


Although tailored to a simple (linear) dynamical setting, the proposed complexity index offers a novel method, grounded in systems theory, for quantifying the complexity that arises from network interactions in dynamical systems.

\noindent\textbf{Organization.} The rest of the paper 
unfolds as follows. In Sec.~\ref{sec:setting}, we describe the class of network systems analyzed in the paper. In Sec.~\ref{sec:background}, we illustrate some background notions and results on McMillan degree and structured systems. In Sec.~\ref{sec:complexity}, we present the proposed complexity index and discuss some of its properties. In Sec.~\ref{sec:network}, we show how the complexity index is related to the topology and the nodal heterogeneity of the network. Sec.~\ref{sec:conclusion} concludes the paper with some remarks and future research directions. 

\noindent\textbf{Notation.} We let $\mathbb{C}$ denote the set of complex numbers, $\mathbb{R}^{n\times m}$ denote the set of $n\times m$ matrices with real entries, and $\mathbb{R}(s)^{n\times m}$ denote the set of $n\times m$ matrix-valued rational functions with real coefficients. We let $[A]_{ij}$ and $A^{\top}$ denote the $(i,j)$-th element and the transpose of matrix $A\in\mathbb{R}^{n\times m}$, respectively. We denote with $\mathrm{diag}(a_{1},a_{2},\dots,a_{n})$ the diagonal matrix with diagonal entries $a_{1}, a_{2},\dots, a_{n}$, and with $\rank(\cdot)$ the rank of a matrix. The symbols $I_{n}$ and $0_{n\times n}$ stand for the $n$-dimensional identity matrix and zero matrix, respectively (we will omit the subscript $n$ when the dimension is clear from the context). For a set $\mathcal{A}$, $|\mathcal{A}|$ denotes its cardinality.

\section{Problem formulation}\label{sec:setting}

We consider a set $\V=\{1,\dots,n\}$ of dynamical agents or nodes. We let $x_i\in\R$ denote the state of the $i$-th node and assume that each state obeys the first-order dynamics
$$
 \dot{x}_i(t) = \gamma_ix_i(t) + u_i(t),\ \ i=1,\dots,n,
$$
where $\gamma_i\in\R$ and $u_{i}(t)$ is a scalar exogenous input.  
Different nodes are allowed to interact with each other according to a directed and weighted interconnection graph $\G=(\V,\E)$, where $\E\subseteq \V\times \V$ is the set of edges of the graph. Such interaction leads to the following collective dynamics:
\begin{align*}
\dot x_i(t) = \gamma_{i}x_i(t) +\!\sum_{\substack{j\,:\,(j,i)\in\E}}\! a_{ij} x_{j}(t)+u_i(t),\ \ i=1,\dots,n
\end{align*}
where $a_{ij}\in\R$ is the weight associated with the edge $(j,i)\in\E$. Letting $A\in\R^{n\times n}$ be the matrix with entries $[A]_{ij}=a_{ij}$ if $(j,i)\in\E$ and $[A]_{ij}=0$ otherwise, and $\Gamma := \diag(\gamma_1,\dots,\gamma_n)$, the collective dynamics above can be written in~vector~form~as
\begin{align}\label{eq:sys}
    \dot x(t) = (\Gamma+A) x(t) + u(t),
\end{align}
where $x(t)=[x_1(t) \cdots\, x_n(t)]^\top$ and $u(t)=[u_1(t) \cdots\, u_n(t)]^\top$ are the vector of states and the vector of inputs, respectively. 
The input-state behavior of \eqref{eq:sys} is described by the matrix-valued transfer function
\begin{align}\label{eq:tf}
    W(s) = (sI-\Gamma-A)^{-1} \in\R(s)^{n\times n}.
\end{align}

The objective of this paper is to introduce and analyze a simple yet principled measure of the complexity of the linear network system \eqref{eq:sys}, designed to explicitly capture the influence of the underlying network structure on the input-output behavior of the system.  
A notable property of our proposed index is that it is almost independent of the particular choice of interconnection weights. More precisely, we shall show that it is a generic property of the system when $A$ is treated as a structured matrix.

\section{Preliminaries}\label{sec:background}

We briefly review some background material on rational matrix-valued transfer functions, structured matrices, and graph theory that are instrumental for the results of the paper.

\subsection{The McMillan degree}



Given a matrix-valued rational function matrix $G(s)\in\R(s)^{q\times m}$, a complex number $\alpha\in\C\cup\{\infty\}$ is called a pole of $G(s)$ if at least one element $[G(s)]_{ij}$ of $G(s)$ has $\alpha$ as a pole. For a pole at infinity ($\alpha=\infty$) this means that $[G(s)]_{ij}\to\infty$ as $s\to\infty$, i.e., $[G(s)]_{ij}$ is not a proper rational function. The degree of a pole of a matrix-valued rational function matrix is defined as follows.

\begin{definition}\emph{ (Degree of a pole)}
Let $G(s)\in\R(s)^{q\times m}$ be a rational matrix-valued function and let $p_i\in\C\cup\{\infty\}$ be a pole of $G(s)$. The degree of $p_i$ is defined as the largest multiplicity that $p_i$ has as a pole over~all~minors~of~$G(s)$.
\end{definition}

Notice that if $G(s)$ is a rational scalar function, the degree of a pole $p_i$ of $G(s)$ is trivially the multiplicity of $p_i$ as a pole of $G(s)$. The notion of zero of a matrix-valued rational function matrix $G(s)\in\R(s)^{q\times m}$ is more subtle. The finite zeros of $G(s)$ can be defined by bringing $G(s)$ to its Smith--McMillan canonical form \cite[Ch.~6]{kailath1980linear}. For the purpose of this paper, it suffices to mention that if $G(s)$ is square and non-singular\footnote{A square matrix-valued rational function $G(s)$ is said to be non-singular, if $\det G(s)$ is not identically zero, i.e., $\det G(s)\neq 0$, for at least one $s\in\C$.} then the zeros of $G(s)$ (including the zero at infinity) coincide with the poles (including the pole at infinity) of $G(s)^{-1}$ \cite[Ch.~6]{kailath1980linear}.

We next recall a key notion for this paper, that is, the McMillan degree of $G(s)\in\R(s)^{q\times m}$.

\begin{definition}\emph{ (McMillan degree)}
Let $\{p_i\}_{i=1}^{n_P}$ be the set of distinct poles of $G(s)\in\R(s)^{q\times m}$ (possibly including
the pole at infinity) and let $\{\delta(G;p_i)\}_{i=1}^{n_P}$ be the corresponding degrees. The McMillan degree of $G(s)$~is~defined~as
$$
\delta_M(G) = \sum\nolimits_{i=1}^{n_P} \delta(G;p_i). 
$$
\end{definition}\smallskip


Loosely speaking, the McMillan degree counts the number of poles of a rational transfer function. For a linear time-invariant system with transfer function $G(s)$, $\delta_M(G)$ equals the smallest dimension of an observable and controllable state-space realization \cite{kalman1965irreducible}. This links the McMillan degree to system complexity: the higher the minimal state dimension, the more complex the system. As such, it has been widely used as a complexity measure in system identification, stochastic realization, and interpolation theory, where reducing or bounding the McMillan~degree~is~often~crucial, see e.g.~\cite{baggio2015minimal, ferrante2012time, georgiou1999interpolation, blomqvist2003matrix}.

While we refer to \cite[Ch.~6]{kailath1980linear} or \cite{youla2015theory} for a complete overview on the properties of the McMillan degree, here we limit to recall some properties which will be instrumental for the results of the paper (for a proof, see e.g.~\cite[Ch.~6.3]{youla2015theory}). 
\begin{proposition}\emph{ (Properties of the McMillan degree)}\label{prop:prop-deltaM}
The\newline following facts hold:
\begin{enumerate}
    \item $\delta_M(G)=0$ if and only if all the entries of $G(s)\in\R(s)^{q\times m}$ are constant real numbers;
    \item $\delta_M(G_1+G_2)\le \delta_M(G_1)+\delta_M(G_2)$, for all $G_1(s)$, $G_2(s)\in\R(s)^{q\times m}$ and the equality holds if and only if $G_1(s)$ and $G_2(s)$ do not share common poles;
    \item $\delta_M(G_1G_2)\le \delta_M(G_1)+\delta_M(G_2)$, for all $G_1(s)\in\R(s)^{q_1\times m}$, $G_2(s)\in\R(s)^{m\times q_2}$. Further, the equality holds if the zeros (poles, resp.) of $G_1(s)$ and the poles (zeros, resp.) of $G_2(s)$ are distinct;
    \item $\delta_M(G^{-1})= \delta_M(G)$ if $G(s)\in\R(s)^{m\times m}$ is square and non-singular.
\end{enumerate}
\end{proposition}

\subsection{Structured matrices and generic properties}

A structured matrix is a matrix in which certain entries are constrained to be zero. Formally, a structured matrix can be described as a binary matrix $A\in\{0,\star\}^{q\times m}$ whose entries are either zero ($0$) or arbitrary real numbers ($\star$). 

A property of a structured matrix $A\in\{0,\star\}^{q\times m}$ is said to be generic if it holds for almost all choices of its free parameters ($\star$). Formally, ``almost all'' means for all free parameters, except those lying on a proper subvariety of the parameter space (typically $\R^N$, where $N$ is the number of free entries of $A$). An alternative probabilistic interpretation is that a property of $A$ is generic if it holds with probability one, if the free parameters are drawn randomly from any continuous distribution. 
When the sparsity pattern of $A\in\{0,\star\}^{q\times m}$ is associated with the interconnection pattern of a graph $\G=(\V,\E)$, then we will say, with a slight abuse of terminology, that a generic property of $A\in\{0,\star\}^{q\times m}$ is a generic property of $\G$.

A matrix property that will be relevant for the results of the paper is the structural rank. \\
\begin{definition}\emph{(Structural rank)}
Let $A\in\{0,\star\}^{q\times m}$ be a structured matrix. The structural rank of $A$, denoted by $\srank(A)$, is the maximum rank of $A$ (in the usual sense) over all numerical realizations of the free parameters of $A$.
\end{definition}

The structural rank is a generic property of $A\in\{0,\star\}^{q\times m}$ in the sense that $\rank(A)$ equals $\srank(A)$ for almost all choices of its free parameters \cite[p.~19]{reinschke1988multivariable}.

The term generic property also refers to properties of dynamical systems described by structured matrices. Well-known examples of such properties are structural controllability and observability of linear systems. For a thorough treatment of structured matrices and generic properties of structured matrices and linear systems described by structured matrices, we refer to the monographs~\cite{reinschke1988multivariable,murota1999matrices}.

\subsection{Graph-theoretic notions}

Consider a directed graph $\G=(\V,\E)$, where $\V=\{1,\dots,n\}$ denotes the set of nodes and $\E\subseteq \V\times \V$ the set of edges. The in-degree of $v\in\V$, denoted by $\deg^-(v)$, is defined as the number of edges in $\E$ pointing into $v$, whereas the out-degree of $v$, denoted by $\deg^+(v)$, is defined as the number of edges in $\E$ pointing out of $v$. A node $v\in\V$ is called source if $\deg^-(v)=0$ and sink if $\deg^+(v)=0$.
An isolated node (i.e., a node $v$ such that $\deg^+(v)=\deg^-(v)=0$) is thus both a source and a sink. The following definitions will be used throughout the paper.


\begin{definition}\emph{ (Maximum matching and matching number)}
A maximum matching of $\G$ is a largest cardinality set of directed edges in which no two edges share a common start node or a common end node. 
The matching number $\nu(\G)$ is the size of a maximum matching.
\end{definition}

It is worth pointing out that there exist efficient, polynomial-time algorithms to compute a maximum matching and matching number of a~directed~graph~\cite[Sec.~D.3]{liu2016control}.

\begin{definition}\emph{ (Ingoing and outgoing edge-cut of a graph)}
Let $\P=\{\P_1,\dots,\P_k\}$, $\V=\cup_{i=1}^k \P_i$, be a partition of $\V$. The ingoing (outgoing) edge-cut of $\G$ induced by $\P$ is the set of subgraphs $\G_i=(\V,\E_i)$, $i=1,\dots,k$, such that each $\E_i\subseteq \E$ contains all the edges in $\E$ pointing into (out of, resp.) the nodes in $\P_i$.
\end{definition}

\section{Quantifying complexity in linear networks}\label{sec:complexity}

\begin{figure}[t!] 
	\centering
\includegraphics[width=0.40\textwidth,trim={0cm 0cm 0cm 0cm},clip]{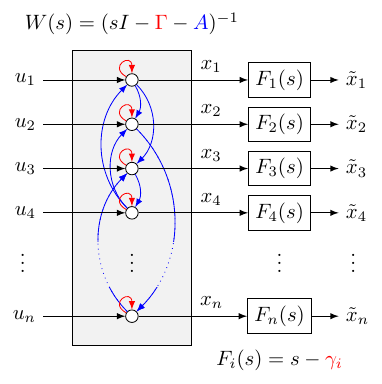}
	\caption{Block-diagram representation of the transfer function $\overline{W}(s)=F(s)W(s)$ that defines the filtered system with ``silenced'' nodal dynamics. We let $F_i(s):=s-\gamma_i$ denote the $i$-th  diagonal entry of $F(s)$.}
    \label{fig:local-deconv}
\end{figure}

To quantify the ``complexity'' of the linear network system \eqref{eq:sys} we look at the McMillan degree of a modified version of its transfer function \eqref{eq:tf}. Specifically, we filter out the local effect of single-node dynamics by applying to the system outputs a local deconvolution, that is, the local (i.e., diagonal) inverse filter
\begin{align*}
    F(s) = sI-\Gamma.
\end{align*}
The transfer function of the resulting filtered system is:
\begin{align}\label{eq:tf2}
\overline{W}(s) := F(s)W(s)
\end{align}
The filtered transfer function $\overline{W}(s)$ can be interpreted as capturing the impact of interconnections on an input reconstruction problem where we aim to recover the global input $u(t)$ based solely on knowledge of isolated node dynamics. The filter $F(s)$ performs a local deconvolution step intended to ``undo'' the known local dynamics (see also Fig.~\ref{fig:local-deconv} for an illustration). What remains in $\overline{W}(s)$ is the residual error dynamics due to the interconnection structure. Note, in particular, that in case of no interconnection between nodes, we have perfect reconstruction and $\overline{W}(s)$ reduces to a static (that is, minimally complex) gain, namely $\overline{W}(s)=I$.


Based on \eqref{eq:tf2}, we propose the following measure of complexity for the network system \eqref{eq:sys}.\\
\begin{definition}\emph{ (Complexity index)}
We define the complexity of the network system \eqref{eq:sys} as
\begin{equation}\label{eq:complexity}
\varphi= \delta_{M}(\overline{W}),
\end{equation}
where $\delta_{M}(\overline{W})\in\mathbb{N}$ denotes the McMillan degree of the transfer function $\overline{W}(s)$ in \eqref{eq:tf2}.
\end{definition}


\begin{remark}\emph{(Stochastic interpretation)}\label{remark:stochastic}
If the inputs $u_{i}(t)$ are i.i.d.~white noise processes with zero mean and unit variance, and the matrix $\Gamma + A$ is Hurwitz stable, then at steady-state the filtered state $\tilde{x}(t)$ in Fig.~\ref{fig:local-deconv} is a stationary stochastic process described by the spectral density
\begin{align}
    \overline{\Phi}(j\omega) & = \overline{W}(j\omega)\overline{W}(-j\omega)^\top, \ \ \omega\in\R. \label{eq:sd}
\end{align}
The spectral density in \eqref{eq:sd} can be analytically extended to the entire complex plane $s\in\C$ (up to a finite number of points), yielding the rational matrix-valued function $\overline{\Phi}(s)=\overline{W}(s)\overline{W}(-s)^\top \in\R(s)^{n\times n}$. In this setting, the complexity index in \eqref{eq:complexity} can be expressed as
$
\varphi = \frac{1}{2}\delta_{M}(\overline{\Phi}),
$
and contains information about the similarity between $\Phi(s):=W(s)W(-s)^{\top}$, the spectral density of $x(t)$, and $\Phi_\ell(s):=F(s)^{-1}F(-s)^{-1}$, the diagonal spectral density of nodal dynamics. In fact, the symmetrized ``ratio'' $$\Phi_2(j\omega)^{-1/2}\Phi_1(j\omega)\Phi_2(-j\omega)^{-1/2}$$
of two spectral densities $\Phi_1(j\omega)$ and $\Phi_2(j\omega)$, where $\Phi_2(j\omega)^{1/2}$ denotes the minimum-phase spectral factor of $\Phi_2(j\omega)$, has been employed to define natural distances between spectral densities, see e.g.~\cite{jiang2012distances}. 
\end{remark}

\begin{lemma}\emph{(Properties of $\varphi$)}\label{lemma:prop}
Consider the network system \eqref{eq:sys} and its complexity index $\varphi$ as defined in \eqref{eq:complexity}. Then, $$0\le \varphi\le n$$ and $\varphi=0$ if and only if $\G$ is totally disconnected, i.e., $\E=\emptyset$.
\end{lemma}
\begin{proof}
The fact that $\varphi\ge 0$ follows immediately from the definition of McMillan degree. To show that $\varphi\le n$, we first note that $\overline{W}(s)$ can be written as
\begin{align}
\overline{W}(s) & =(sI-\Gamma)(sI-\Gamma-A)^{-1} \notag\\
& = ((sI-\Gamma-A)(sI-\Gamma)^{-1})^{-1}\notag \\
& = (I-A(sI-\Gamma)^{-1})^{-1}, \label{eq:equiv}
\end{align}
so that, from the properties of the McMillan degree (Proposition \ref{prop:prop-deltaM}), we have
\begin{align}
\varphi & =  \delta_{M}(\overline{W}) = \delta_{M}\left((I-A(sI-\Gamma)^{-1})^{-1}\right) 
\nonumber \\
& = \delta_{M}\left(I-A(sI-\Gamma)^{-1}\right) \label{eq:prooflem2} \\
& = \delta_{M}\left(A(sI-\Gamma)^{-1}\right) \label{eq:prooflem3} \\
& \le \underbrace{\delta_{M}(A)}_{=0} + \underbrace{\delta_{M}\left((sI-\Gamma)^{-1}\right)}_{=n}= n, \label{eq:prooflem4}
\end{align}
where \eqref{eq:prooflem2} follows from the fact that $\overline{W}(s)=(I-A(sI-\Gamma)^{-1})^{-1}$ is non-singular and property 4) of Proposition \ref{prop:prop-deltaM}, \eqref{eq:prooflem3} follows from property 2) of Proposition \ref{prop:prop-deltaM}, and \eqref{eq:prooflem4} from property 3) of Proposition \ref{prop:prop-deltaM}. 
Finally, if $\G$ is totally disconnected then $A=0_{n\times n}$ and $\overline{W}(s)=I$, which implies $\varphi=0$. Conversely, if $\G$ is not totally disconnected ($\E\ne\emptyset$), then $A$ has at least one non-zero (off-diagonal) entry. Suppose, without loss of generality, that $A$ has a non-zero entry in position $(i,j)$, $i\ne j$. This implies that $A(sI-\Gamma)^{-1}$ has at least one entry featuring a (finite) pole,~and~from~\eqref{eq:equiv}
\begin{align*}
    [\overline{W}(s)^{-1}]_{ij} & =-a_{ij}/(s-\gamma_i).
\end{align*}
Thus, $\overline{W}(s)^{-1}$ features a (finite) pole and, from property 4) of Proposition \ref{prop:prop-deltaM}, $\delta_{M}(\overline{W})=\delta_{M}(\overline{W}^{-1})>0$, which implies that $\varphi>0$. 
\end{proof}

In the following definition, we extend the previous notion of complexity to make it independent of the specific values of the weights of the interconnection graph $\G$.

\begin{definition}\emph{ (Structural complexity index)}
We define the structural complexity of the network system \eqref{eq:sys}, denoted by $\varphi_{\G}$, as the maximum complexity $\varphi$ of \eqref{eq:sys} over all numerical realizations of the edge weights of $\G$.
\end{definition}

The complexity index $\varphi$ of the network system \eqref{eq:sys} is generically equal to $\varphi_\G$, as we show next.
\begin{proposition}\emph{(Complexity is a generic property)}\label{prop:genericity}
Consider the network system \eqref{eq:sys}, where $A$ is interpreted as the structured matrix with sparsity pattern described by the graph $\G$. Then $\varphi=\varphi_{\G}$ for almost all choices of the weights of $\G$.
\end{proposition}
\begin{proof}
Consider the transfer function 
\begin{align}\label{eq:Q}
    Q(s) = A(sI-\Gamma)^{-1},
\end{align}
From Eq.~\eqref{eq:prooflem3}, it holds $\delta_M(\overline{W}) = \delta_M(Q)$. Note that $A$ is a constant matrix and $(sI-\Gamma)^{-1}$ is diagonal, so that all the poles of $Q(s)$ are finite and are a subset of the diagonal entries of $\Gamma$. Assume that $\{\gamma^*_{i}\}_{i=1}^k\subseteq\{\gamma_i\}_{i=1}^n$ are the distinct poles of $Q(s)$ and note that we can express $Q(s)$ as a sum of matrix terms featuring distinct poles as
\begin{align}\label{eq:QQ}
    Q(s) = \sum\nolimits_{i=1}^k C_i / (s-\gamma^*_i)
\end{align}
where $C_i$ are $n\times n$ constant matrices whose non-zero columns are also columns of $A$. Let $Q_i(s):=C_i/(s-\gamma^*_i)$. From the definition of McMillan degree and property 2) of Proposition \ref{prop:prop-deltaM}, it holds
\begin{align}\label{eq:deltaMlemma}
    \delta_M(Q) = \sum\nolimits_{i=1}^k \delta_M(Q_i) = \sum\nolimits_{i=1}^k \delta\left(Q_i;\gamma_i^*\right).
\end{align}
The degree of the pole $\gamma_i^*$ of $Q_i$, i.e.,~$\delta\left(Q_i;\gamma_i^*\right)$, is by definition the maximum multiplicity of the pole $\gamma_i^*$ over all minors of $Q_i$. Owing to the special structure of $Q_i(s)$, this implies that
$$
\delta\left(Q_i;\gamma_i\right) = \rank(C_i)
$$ 
because the rank of a matrix is the size of the largest non-zero minor of the matrix. Since the structural rank of $A$ (i.e.~the maximum rank over all numerical realizations of $A$) is a generic property of a structured matrix \cite[p.~19]{reinschke1988multivariable}, each $\delta\left(Q_i;\gamma_i^*\right)$ takes the same value for almost all choices of the non-zero entries of $A$. Note that this value is equal to the maximum of $\rank(C_i)$ over all numerical realizations of $A$, i.e., $\srank(C_i)$. Finally, from \eqref{eq:deltaMlemma} and the identity $\varphi=\delta_M(\overline{W})=\delta_M(Q)$, we conclude that $\varphi$ takes the same value $\varphi_{\mathcal{G}}$ for almost all choices of the non-zero entries~of~$A$.~
\end{proof}

It is worth highlighting that a key step in the proof of Proposition \ref{prop:genericity} is the characterization of $\varphi$ in terms of the matrix $Q(s)$ in \eqref{eq:Q}. This allows, through the decomposition in \eqref{eq:QQ}, to rewrite the proposed complexity index as a sum of ranks of specific submatrices of $A$.

\section{Complexity and network structure}\label{sec:network}

The next result sheds light on the relationship between the structural complexity index $\varphi_{\mathcal{G}}$, the topology of $\G$, and the pattern of heterogeneity of nodal dynamics.

\begin{theorem}\emph{ (Link between $\varphi_{\mathcal{G}}$ and network structure)}\label{thm:link-complexity-network}
Consider the network system \eqref{eq:sys}. Let $\{\gamma_i^*\}_{i=1}^{k}\subseteq\{\gamma_i\}_{i=1}^{n}$ denote the distinct values in the set $\{\gamma_i\}_{i=1}^{n}$ and let 
\begin{align*}
\P_i := \{j\in\V : \gamma_j = \gamma_i^*\}, \quad i=1,\dots,k.
\end{align*}
Then
\begin{align*}
    \varphi_{\G} = \sum\nolimits_{i=1}^k \nu(\G_i),
\end{align*}
where $\{\G_i\}_{i=1}^k$ is the outgoing edge-cut of $\G$ induced by $\P=\{\P_1,\dots,\P_k\}$ and $\nu(\G_i)$ denotes the matching number of the subgraph $\G_i$ of $\G$.
\end{theorem}
\begin{proof}
Following the same arguments of the proof of Proposition \ref{prop:genericity}, it holds
\begin{align*}
    \varphi_{\mathcal{G}} = \sum\nolimits_{i=1}^k \srank(C_i),
\end{align*}
where the entries $[C_i]_{jh}$ of the $n\times n$ constant matrices $C_i$, $i=1,\dots,k$, are given by
\begin{align}\label{eq:Ci}
[C_i]_{jh} = \begin{cases} [A]_{jh}, &\text{if }h\in\P_i, \\ 0, &\text{otherwise.} \end{cases} 
\end{align}
Observe that $C_i^\top$ coincides with the adjacency matrix of the (weighted and directed) subgraph $\G_i$.
The result now follows from the facts that (i) the rank of a matrix equals the rank of its transpose, and (ii) the structural rank of a structured matrix $M$ equals the matching number of the directed graph described by the adjacency matrix $M$, see e.g.~\cite[Sec.~2.1.3]{murota1999matrices}.   
\end{proof}

Theorem \ref{thm:link-complexity-network} reveals that the structural complexity index $\varphi_{\mathcal{G}}$ depends on the partitioning of nodes into groups with identical dynamics 
and, precisely, on the interconnection structure of the subgraphs $\{\mathcal{G}_i\}_{i=1}^k$ induced by such partition, as captured by their matching number. 
Intuitively, greater heterogeneity in local dynamics (in terms of number of groups with different nodal dynamics) leads to a higher $\varphi_{\mathcal{G}}$, while an increased number of independent edges (i.e.~not sharing start or end nodes) of $\mathcal{G}$ also results in~a~higher~$\varphi_{\mathcal{G}}$.~
\begin{remark}\emph{(Group-wise homogeneity)} In real network systems, the internal dynamics described by $\gamma_i$'s typically differ from node to node. However, it is often reasonable to group nodes with similar dynamical properties and model them as having identical local dynamics. 
For instance, neurons can be grouped according to their physiological characteristics, individuals in social networks by their level of stubbornness or bias towards certain opinions, and components in a power grid by their design specifications.
\end{remark}

\begin{figure}[t!] 
	\centering
	\subfigure[Scale-free network]{\includegraphics[width=0.43\textwidth,trim={0cm 0cm 0cm 0cm},clip]{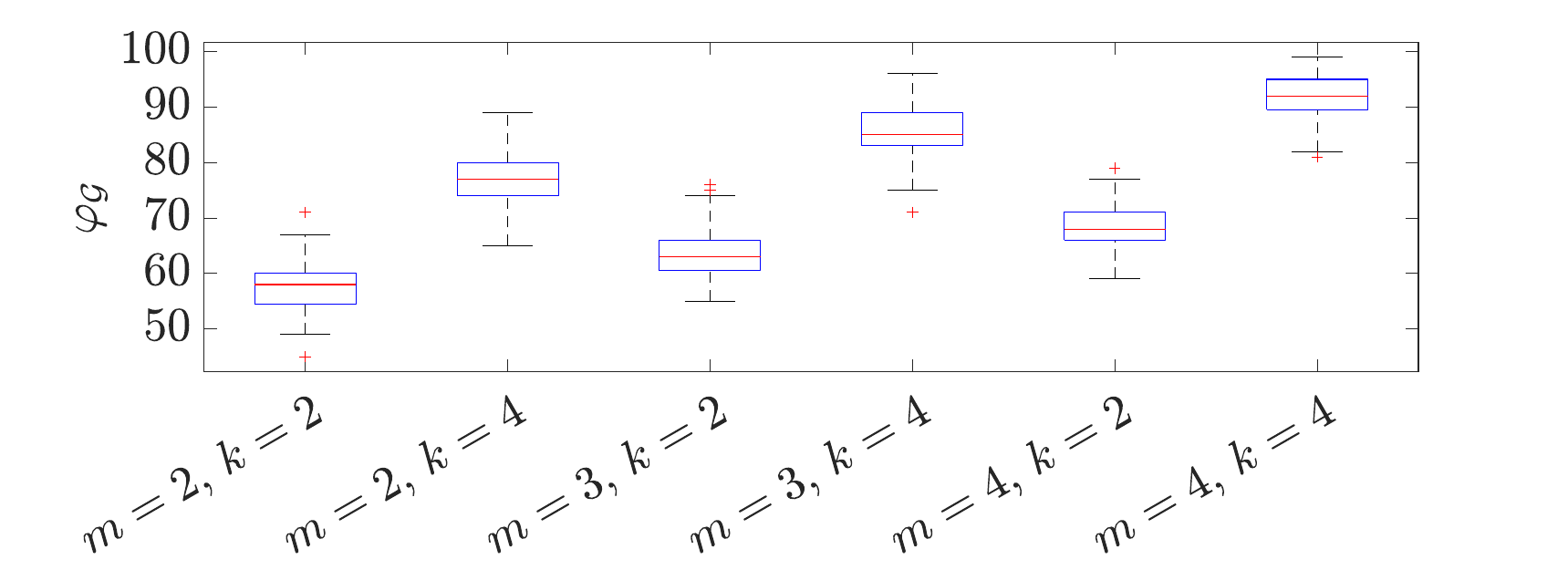}\label{fig:scalefree}} \\
	\subfigure[Small-world network]{\includegraphics[width=0.43\textwidth,trim={0cm 0cm 0cm 0cm},clip]{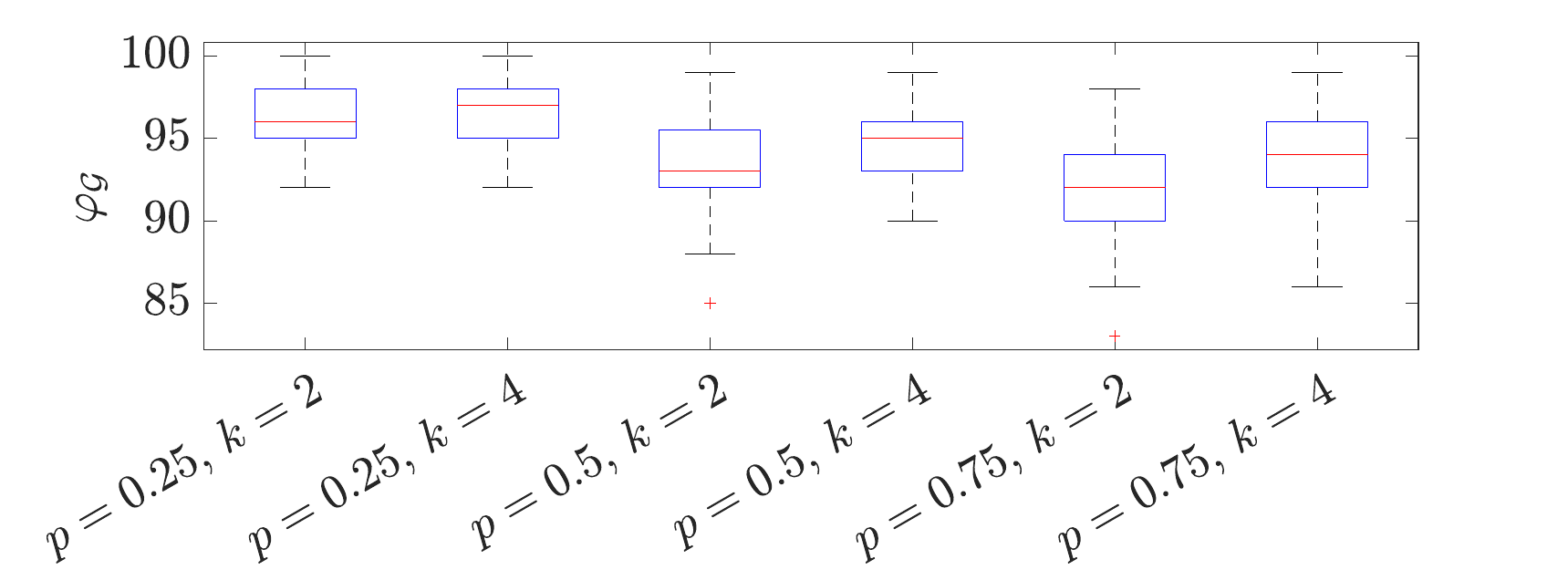}\label{fig:smallworld}}
	\caption{Structural complexity of two types of random networks, each with $n=100$ nodes and $k$ randomly assigned groups exhibiting distinct dynamics. (a) Barab\'asi--Albert scale-free networks with preferential attachment parameter $m$. (b) Watts--Strogatz small-world networks with rewiring probability $p$. The box plots show the outcomes of $N=100$~Monte~Carlo~runs.}
	\label{fig:num_res}
\end{figure}

In Fig.~\ref{fig:num_res}, we illustrate the complexity $\varphi_{\mathcal{G}}$ of two types of random networks, namely scale-free networks generated using the Barab\'asi--Albert model \cite{barabasi1999emergence} and small-world networks generated via the Watts--Strogatz model \cite{watts1998collective}. Both networks consist of $n=100$ nodes, divided into $k$ randomly chosen groups with distinct nodal dynamics. In both cases, the complexity $\varphi_{\mathcal{G}}$ increases with $k$. For scale-free networks, $\varphi_{\mathcal{G}}$ grows with the preferential attachment parameter $m$, which controls the number of edges added at each step of the network generation process. In small-world networks, the total number of edges remains constant as the rewiring probability $p$ varies. However, higher values of $p$ lead to lower complexity due to the presence~of~more~node-sharing~edges.

\begin{table}[t]
	\centering
	\caption{\normalfont \parbox{\linewidth}{\medskip Normalized complexity index $\varphi_{\mathcal{G}}$ for three real network systems. The \emph{C.~Elegans} network (CE) represents the chemical synapse connectome of the nematode \emph{C.~Elegans}, as described in~\cite{varshney2011structural}. The \emph{Northern European Power Grid} (PG) refers to the power transmission network analyzed in~\cite{menck2014dead}. The \emph{US Political Blogs} network (PB) corresponds to the directed hyperlink network between political weblogs in the United States, documented in~\cite{adamic2005political}. For each network, we adopt the following groupings of internal dynamics (available in the above references): sensory, interneurons, and motor neurons for CE; net generators and net consumers for PG; and liberal and conservative blogs for PB. Values under ``true'' refer to the original networks, while ``rand'' indicates randomized versions obtained by rewiring edges between node pairs selected uniformly at random, keeping the total number of edges unchanged. For the randomized case, the results show the mean over 100 Monte Carlo runs.}}
	\label{tab:comparison_real_nets}
	\setlength{\tabcolsep}{4pt}
	\begin{tabular}{lcccc}
		\toprule
		\textbf{ Network type} & { $n$} & { $\varphi_{\mathcal{G}}/n$  (true)} & { $\varphi_{\mathcal{G}}/n$ (rand)}  \\
		\midrule
		{ CE } &  $279$        & $0.9032$   & $0.9996$      \\
		{ PG } & $236$   & $0.8644$  & $0.7077$   \\
		{ PB } & $1490$            & $0.6148$    & $1$    \\
		\bottomrule
	\end{tabular}
\end{table}
In Table \ref{tab:comparison_real_nets}, we compute the index $\varphi_{\mathcal{G}}$ normalized by network size for three real-world network systems from different domains and compare it with versions of these networks where the edges have been rewired uniformly at random\footnote{The code used in these experiments is available on GitHub at:\\ {\scriptsize \url{https://github.com/marcofabris92/network-complexity}}.}. The gap between true and randomized values (particularly evident for the \emph{US Political Blogs} network) suggests that the index captures some meaningful structural information of real network systems, while the higher value observed in the \emph{C.~Elegans} network supports the idea that more functionally advanced systems tend~to~exhibit~greater~complexity.

We conclude this section by presenting a corollary of Theorem \ref{thm:link-complexity-network} which provides simple bounds on $\varphi_{\mathcal{G}}$ not depending on partitions induced by nodal dynamics. 

\begin{corollary}\label{cor:simpleboundsonphiG}{\em (Simple bounds on $\varphi_{\mathcal{G}}$)}
Consider the network system \eqref{eq:sys}. Then, 
\begin{align*}
     \quad\quad \nu(\G)\le \varphi_{\G} \le n- |S(\G)|,
\end{align*}
where $\nu(\G)$ and $S(\G)$ denote the matching number and the set of sink nodes of $\mathcal{G}$, respectively.
Moreover, the first inequality is attained when all $\{\gamma_i\}_{i=1}^n$ are identical, while the second inequality when all $\{\gamma_i\}_{i=1}^n$ are distinct.
\end{corollary}
\begin{proof}
Let $C_i$, $i=1,\dots,k$, be defined as in \eqref{eq:Ci} and let $A_j$, $j=1,\dots,n$, be $n\times n$ zero matrices except for their $i$-th column which equals the $i$-th column of $A$. It holds
$$
\rank(A) \le \sum\nolimits_{i=1}^k\rank(C_i)\le \sum\nolimits_{j=1}^n\rank(A_j),
$$
since no column of $A$ appears in more than one $C_i$. Note that when all $\{\gamma_i\}_{i=1}^n$ are identical $k=1$ and $C_1=A$ and when all $\{\gamma_i\}_{i=1}^n$ are distinct $k=n$ and $C_i=A_i$, $i=1,\dots,n$.
When $A$ is interpreted as a structured matrix the rank in the previous inequalities can be replaced by the structural rank, and the result follows by noting that $\srank(A)=\nu(\G)$ and $\sum_{j=1}^n\srank(A_j)= n- |S(\G)|$, where the latter identity holds because $A_j=0_{n\times n}$~if~and~only~if~$j\in S(\G)$.~
\end{proof}

\begin{remark}\emph{(Relation with structural controllability)}
	Consider a network system \eqref{eq:sys} with identical nodal dynamics, i.e.,~$\gamma_i = \gamma_j$ for all $i,j$. In this case, the minimum number of inputs $N_{\text{min}}$ needed to guarantee structural controllability~is~given~by \cite{liu2016control}
	\begin{align*}
		N_{\text{min}} = \max\{1, n - \nu(\mathcal{G})\}.
	\end{align*}
	This reveals a complementary relation between $\varphi_{\mathcal{G}}$ and $N_{\text{min}}$: richer internal dynamics facilitate signal propagation, reducing the need for inputs. 
\end{remark}

\section{Conclusions and future work}\label{sec:conclusion}

We have introduced a network complexity index tailored to systems composed of interconnected linear first-order dynamical nodes. Defined in terms of the McMillan degree of the system after locally filtering out nodal dynamics, our index aims at capturing the influence of interconnections on overall system complexity. We show that this index is a generic property of the network system, yielding the same values for almost all choices of interconnection weights. Additionally, we provide a graph-theoretic characterization, clarifying how the interconnection pattern and nodal dynamics affect the proposed complexity index.

The driving idea behind this work is that when measuring network complexity in interconnected systems the dynamic nature of the system cannot be neglected. 
A limitation of our index is that it is tailored to first-order scalar nodal dynamics. While we believe that some results can be extended to certain classes of higher-order network systems (particularly those with interconnections through zeroth-order, position-like variables) the general case requires further investigation. Another limitation of our index lies in its discrete nature. It would be interesting to explore real-valued complexity metrics that account for edge weights, such as distances between the spectral densities of systems described by isolated and interconnected nodes driven by noise (see also Remark~\ref{remark:stochastic}). Additionally, it would be interesting to investigate the role played by structural constraints, such as those imposed by Laplacian dynamics, on our index.
More generally, we hope that this work will inspire the development of new measures that more effectively capture the contribution of the network to the complexity of~interconnected~dynamical~systems.\smallskip

\bibliographystyle{abbrv}
\bibliography{ref_final}

\begin{thebibliography}{10}

\bibitem{adamic2005political}
L.~A. Adamic and N.~Glance.
\newblock The political blogosphere and the 2004 us election: divided they
  blog.
\newblock In {\em Proceedings of the 3rd International Workshop on Link
  Discovery}, pages 36--43, 2005.

\bibitem{baggio2015minimal}
G.~Baggio and A.~Ferrante.
\newblock On minimal spectral factors with zeroes and poles lying on prescribed
  regions.
\newblock {\em IEEE Trans. on Automatic Control}, 61(8):2251--2255, 2015.

\bibitem{barabasi1999emergence}
A.-L. Barab{\'a}si and R.~Albert.
\newblock Emergence of scaling in random networks.
\newblock {\em Science}, 286(5439):509--512, 1999.

\bibitem{blomqvist2003matrix}
A.~Blomqvist et~al.
\newblock Matrix-valued {N}evanlinna-{P}ick interpolation with complexity
  constraint: {A}n optimization approach.
\newblock {\em IEEE Trans. on Automatic Control}, 48(12):2172--2190, 2003.

\bibitem{bonchev2005quantitative}
D.~Bonchev and G.~A. Buck.
\newblock Quantitative measures of network complexity.
\newblock In {\em Complexity in chemistry, biology, and ecology}, pages
  191--235. Springer, 2005.

\bibitem{carlson2002complexity}
J.~M. Carlson and J.~Doyle.
\newblock Complexity and robustness.
\newblock {\em Proceedings of the national academy of sciences},
  99(suppl\_1):2538--2545, 2002.

\bibitem{claussen2007offdiagonal}
J.~C. Claussen.
\newblock Offdiagonal complexity: A computationally quick complexity measure
  for graphs and networks.
\newblock {\em Physica A: Statistical Mechanics and its Applications},
  375(1):365--373, 2007.

\bibitem{costa2007characterization}
L.~d.~F. Costa et~al.
\newblock Characterization of complex networks: A survey of measurements.
\newblock {\em Advances in physics}, 56(1):167--242, 2007.

\bibitem{ferrante2012time}
A.~Ferrante, C.~Masiero, and M.~Pavon.
\newblock Time and spectral domain relative entropy: {A} new approach to
  multivariate spectral estimation.
\newblock {\em IEEE Trans. on Automatic Control}, 57(10):2561--2575, 2012.

\bibitem{georgiou1999interpolation}
T.~T. Georgiou.
\newblock The interpolation problem with a degree constraint.
\newblock {\em IEEE Trans. on Automatic Control}, 44(3):631--635, 1999.

\bibitem{grone1988bound}
R.~Grone and R.~Merris.
\newblock A bound for the complexity of a simple graph.
\newblock {\em Discrete mathematics}, 69(1):97--99, 1988.

\bibitem{harrison2016role}
W.~K. Harrison.
\newblock The role of graph theory in system of systems engineering.
\newblock {\em IEEE Access}, 4:1716--1742, 2016.

\bibitem{jiang2012distances}
X.~Jiang, L.~Ning, and T.~T. Georgiou.
\newblock Distances and riemannian metrics for multivariate spectral densities.
\newblock {\em IEEE Trans. on Automatic Control}, 57(7):1723--1735, 2012.

\bibitem{kailath1980linear}
T.~Kailath.
\newblock {\em Linear systems}, volume 156.
\newblock Prentice-Hall Englewood Cliffs, NJ, 1980.

\bibitem{kalman1965irreducible}
R.~Kalman.
\newblock Irreducible realizations and the degree of a rational matrix.
\newblock {\em J. of the Society for Ind. and Appl. Mathematics},
  13(2):520--544, 1965.

\bibitem{kim2008complex}
J.~Kim and T.~Wilhelm.
\newblock What is a complex graph?
\newblock {\em Physica A: Statistical Mechanics and its Applications},
  387(11):2637--2652, 2008.

\bibitem{ladyman2013complex}
J.~Ladyman, J.~Lambert, and K.~Wiesner.
\newblock What is a complex system?
\newblock {\em European Journal for Philosophy of Science}, 3:33--67, 2013.

\bibitem{liu2016control}
Y.-Y. Liu and A.-L. Barab{\'a}si.
\newblock Control principles of complex systems.
\newblock {\em Reviews of Modern Physics}, 88(3):035006, 2016.

\bibitem{lloyd2001measures}
S.~Lloyd.
\newblock Measures of complexity: a nonexhaustive list.
\newblock {\em IEEE Control Systems Magazine}, 21(4):7--8, 2001.

\bibitem{menck2014dead}
P.~J. Menck, J.~Heitzig, et~al.
\newblock How dead ends undermine power grid stability.
\newblock {\em Nature communications}, 5(1):3969, 2014.

\bibitem{minoli1975combinatorial}
D.~Minoli.
\newblock Combinatorial graph complexity.
\newblock {\em Atti della Acc. Naz. dei Lincei. Cl. di Scienze Fis., Mat. e
  Nat.. Rendiconti}, 59:651--661, 1975.

\bibitem{murota1999matrices}
K.~Murota.
\newblock {\em Matrices and matroids for systems analysis}, volume~20.
\newblock Springer Science \& Business Media, 1999.

\bibitem{neel2013linear}
D.~L. Neel and M.~E. Orrison.
\newblock The linear complexity of a graph.
\newblock {\em Advances in Network Complexity}, pages 155--175, 2013.

\bibitem{reinschke1988multivariable}
K.~J. Reinschke.
\newblock Multivariable control a graph theoretic approach.
\newblock {\em Lecture notes in control and information sciences}, 108, 1988.

\bibitem{varshney2011structural}
L.~R. Varshney, B.~L. Chen, E.~Paniagua, D.~H. Hall, and D.~B. Chklovskii.
\newblock Structural properties of the caenorhabditis elegans neuronal network.
\newblock {\em PLoS computational biology}, 7(2):e1001066, 2011.

\bibitem{watts1998collective}
D.~J. Watts and S.~H. Strogatz.
\newblock Collective dynamics of ‘small-world’ networks.
\newblock {\em nature}, 393(6684):440--442, 1998.

\bibitem{youla2015theory}
D.~C. Youla.
\newblock {\em Theory and synthesis of linear passive time-invariant networks}.
\newblock Cambridge University Press, 2015.

\end{thebibliography}

\end{document}